\documentclass[pra,twocolumn,superscriptaddress]{revtex4}
\usepackage{psfrag}
\usepackage{amssymb}
\usepackage[dvips]{graphicx}
\usepackage[latin1]{inputenc}
\usepackage{color}

\begin{document}

\title{Interference effects in above-threshold ionization from diatomic
molecules: determining the internuclear separation}
\author{H. Hetzheim}
\affiliation{Max-Planck-Institut f\"ur Kernphysik, Saupfercheckweg 1, 69117 Heidelberg,
Germany}
\affiliation{Max-Born-Institut f\"ur nichtlineare Optik und Kurzzeitspektroskopie,
Max-Born-Str. 2A, D-12489 Berlin, Germany}
\author{C. Figueira de Morisson Faria}
\affiliation{Department of Physics and Astronomy, University College London, Gower
Street, London WC1E 6BT, United Kingdom}
\author{W. Becker}
\affiliation{Max-Born-Institut f\"ur nichtlineare Optik und Kurzzeitspektroskopie,
Max-Born-Str. 2A, D-12489 Berlin, Germany}
\date{\today}

\begin{abstract}
We calculate angle-resolved above-threshold ionization spectra for diatomic
molecules in linearly polarized laser fields, employing the strong-field
approximation. The interference structure resulting from the individual
contributions of the different scattering scenarios is discussed in detail,
with respect to the dependence on the internuclear distance and molecular
orientation. We show that, in general, the contributions from the processes
in which the electron is freed at one center and rescatters off the other
obscure the interference maxima and minima obtained from single-center
processes. However, around the boundary of the energy regions for which
rescattering has a classical counterpart, such processes play a negligible
role and very clear interference patterns are observed. In such energy
regions, one is able to infer the internuclear distance from the energy
difference between adjacent interference minima.
\end{abstract}

\maketitle

\section{\label{sec:level1} Introduction}

The interaction of matter with an intense laser field ($I\gtrsim 10^{13}%
\mathrm{W/cm}^{2}$) leads to several phenomena, such as above-threshold
ionization (ATI) or high-order harmonic generation (HHG). Such phenomena owe
their existence to physical mechanisms, in which an electron reaches the
continuum, by tunneling or multiphoton ionization, at an instant $t^{\prime
} $. Subsequently, it is accelerated by the field and driven back towards
its parent ion, or molecule, with which it rescatters or recombines at a
later time $t$ \cite{tstep}. Such laser-induced recombination or
rescattering processes take place within a fraction of a laser-field cycle.
The period of a typical near-infrared Ti:sapphire laser pulse is $T=2\pi
/\omega \sim 2.6 \mathrm{fs}$. Thus, HHG and ATI occur on a time scale of
hundreds of attoseconds \cite{Scrinzi2006}. Hence, above-threshold
ionization and high-order harmonic generation may be employed for probing,
or even controlling, dynamic processes with attosecond and sub-angstrom
resolution.

This fact, together with new alignment techniques, has opened a whole new
range of possibilities for studying molecules in strong laser fields,
employing high-energy photoelectrons or high-order harmonic radiation.
Concrete examples are the attosecond reconstruction of the nuclear motion in
a molecule \cite{Baker2006}, the real-time imaging of vibrational
wavepackets \cite{vibration}, the tomographic reconstruction of molecular
orbitals \cite{Itatani2004}, the time-resolved measurement of intramolecular
quantum-interference effects \cite{tokyo2005}, or the determination of
internuclear distances \cite{Hu2005}.

These applications are a direct consequence of the fact that a molecule
possesses a very specific configuration of ions from which the electron may
leave, or off which it may rescatter causing above-threshold ionization, or
recombine generating high-harmonics. This leads to characteristic
quantum-interference patterns in the HHG or ATI spectra, in which structural
information about the molecule is hidden. This is true both for polyatomic
\cite{organic} and diatomic molecules \cite%
{interfexp,tokyo2005,Hu2005,MBBF00,doubleslit,KBK98,ATIangleexp,Madsen,Usach2006,Usachenko,Kansas,Hetzheim2005,PRACL2006,DM2006,SFAvibrMads,SFAvibrLein,SFAvibrBeck,fewcyc,elliptical}%
. In particular for diatomic molecules, it has been shown that the
high-order harmonic or ATI spectra exhibit overall maxima and minima, which
are highly dependent on the spatial separation between both centers in the
molecules, and can be described as the interference between two radiating
point sources. In this sense, HHG or ATI by a diatomic molecule may be
viewed as the microscopic analog of a double-slit experiment \cite%
{tokyo2005,MBBF00,doubleslit}. Furthermore, such features depend on the
symmetry of the highest occupied molecular orbitals, and on the alignment
angle of the molecule with respect to the laser-field polarization \cite%
{tokyo2005,Hu2005,MBBF00,doubleslit,KB2005,KBK98,Madsen,Usach2006,Usachenko,Kansas,Hetzheim2005,PRACL2006,DM2006,SFAvibrMads,SFAvibrLein,SFAvibrBeck}%
.

Specifically in the diatomic case, several aspects of this interference have
been extensively studied in the past few years, such as the influence of the
orbital symmetry, the internuclear distance, the alignment angles \cite%
{tokyo2005,Hu2005,MBBF00,doubleslit,KB2005,KBK98,Madsen,Usach2006,Usachenko,Kansas,Hetzheim2005,PRACL2006,DM2006}%
, and molecular vibration \cite{SFAvibrMads,SFAvibrLein,SFAvibrBeck}, as
well as the role of the laser-field shape \cite{fewcyc,fewcycvibr} or polarization \cite%
{elliptical}. Furthermore, an adequate modeling of bound molecular states,
in comparison with existing ionization experiments \cite{ATIangleexp}, has
also raised considerable debate \cite{Usachenko,Usach2006,Madsen,DM2006}.

For that purpose, both the purely numerical solution of the time-dependent
Schrödinger equation \cite{Hu2005,doubleslit,KB2005}, and the strong-field
approximation \cite%
{KBK98,MBBF00,Madsen,Usach2006,Usachenko,Kansas,PRACL2006,DM2006,SFAvibrMads,SFAvibrLein,SFAvibrBeck}
have been employed. The latter method allows a transparent physical
interpretation of the phenomena in question as laser-induced rescattering or
recombination processes, and permits a clear space-time picture, which can
be related to the classical orbits of an electron in a strong laser field
\cite{orbitshhg}. For a diatomic molecule, there exist two main rescattering
or recombination scenarios: the electron born through ionization at the
center $C_{i}$ upon its return may recollide and interact with either the
same ion $(C_{i})$, or with the other one ( $C_{j}$ $(i\neq j)$) Such
processes have been taken into account for high-order harmonic generation
employing a two-center zero-range potential \cite{KBK98,Hetzheim2005}, using
Bessel function expansions \cite{Usach2006}, and by means of saddle-point
methods \cite{Hetzheim2005,PRACL2006}.

In this paper, we calculate the energy spectra and angular distributions of
ATI produced electrons in linearly polarized laser fields, within the
framework of the strong-field approximation (SFA) and the single-active
electron approximation (SAE). We employ a zero-range potential model similar
to that in \cite{KBK98}, and consider both the direct electrons, which reach
the detector without interacting with their parent molecule, and the
electrons that suffer a single act of rescattering before reaching the
detector. In the latter case, we put particular emphasis on interference
effects: A final state with given momentum outside the laser field can be
reached via two different scenarios. An electron can be born at and
rescatter off the same center, or it can be born at one center and rescatter
off the other. We show that the processes involving two centers, in general,
obscure the interference patterns in the ATI spectra, in almost all
energy-angle regions. An exception, however, is the boundary of the region
that after tunneling is classically accessible to the ionized electron, in
other words, the region before the classical cutoff. Near this boundary, the
two-center processes yield negligible contributions, and one may identify
very clear interference patterns. This makes it possible to provide a recipe
to determine the internuclear distance $R$ out of the angle-resolved ATI
spectra. Throughout the article we will use the velocity gauge and atomic
units $(e=m=\hbar=1,\ c=137)$.

The paper is organized as follows: In Sec. \ref{transampl} we provide the
ATI transition amplitudes for the direct and for the rescattered electrons,
which in Sec. \ref{results} are employed to compute the ATI spectra. The
interference patterns in the spectra are analyzed with respect to molecular
orientation, internuclear distance, and the position of the detector with
respect to the polarization of the laser field and the molecular axis (Sec %
\ref{alignment}). In Sec. \ref{contourplots}, we present angle-resolved
spectra, from which we infer the internuclear distance. Finally, in Sec. \ref%
{conclusions}, we summarize the paper.

\section{Transition amplitudes}

\label{transampl}

The transition amplitude for direct ionization, within the strong-field
approximation (SFA) \cite{footnote}, is given by the Keldysh-Faisal-Reiss
amplitude \cite{KelFaisReiss}
\begin{equation}
M_{\mathbf{p}}=-i\int_{-\infty }^{\infty }dt\langle \Psi _{\mathbf{p}%
}^{(V)}(t)|V|\Psi _{0}(t)\rangle,  \label{Eq.:KFR}
\end{equation}%
where $|\Psi _{0}(t)\rangle =|\Psi _{0}\rangle \exp(i|E_{0}|t)$. The
amplitude describes an electron, initially in the ground state $|\Psi
_{0}\rangle $, that is injected in the continuum by the laser field
overcoming the ionization potential $|E_{0}|$, and reaches the detector with
final momentum $\mathbf{p}$. The form of the transition amplitude given
here, which contains the binding potential $V(\mathrm{r})$ rather than the
interaction with the laser field, was first presented in Ref.~\cite{PPT}. In
the SFA, the final state with momentum $\mathbf{p}$ is described by a Volkov
state, which in velocity gauge has the form
\begin{equation}
\langle \mathbf{r}|\Psi _{\mathbf{p}}^{(V)}(t)\rangle =\frac{1}{(2\pi )^{3/2}%
}e^{i\mathbf{p}\cdot \mathbf{r}}e^{-\frac{i}{2}\int_{-\infty }^{t}d\tau
\lbrack \mathbf{p}+\mathbf{A}(\tau )]^{2}}.  \label{Eq.:Volkov}
\end{equation}%
In the amplitude (\ref{Eq.:KFR}), the electron once ionized does not
interact with the ion (that is, with the binding potential $V$) anymore. If
we allow for at most one single act of rescattering, the amplitude (\ref%
{Eq.:KFR}) is replaced by
\begin{eqnarray}
M_{\mathbf{p}}^{(0,1)} &=&-\int_{-\infty }^{\infty }dt\int_{-\infty
}^{t}dt^{\prime }  \nonumber \\
&&\times \langle \Psi _{p}^{(V)}(t)|VU^{(V)}(t,t^{\prime })V|\Psi
_{0}(t^{\prime })\rangle .  \label{Eq.:rescatter}
\end{eqnarray}%
Here, $U^{(V)}(t,t^{\prime })$ denotes the Volkov time-evolution operator,
which describes the evolution of the electron in the presence of the
external laser field, ignoring the binding potential. Equation~(\ref%
{Eq.:rescatter}) incorporates direct ionization, as described by Eq.~(\ref%
{Eq.:KFR}), as well as ionization followed by rescattering (for details,
see, e.g., Ref.~\cite{LKKB97}).

In order to apply Eqs.~(\ref{Eq.:KFR}) and (\ref{Eq.:rescatter}) to a
diatomic molecule, we consider the two-center binding potential
\begin{equation}
V(\mathbf{r})=V_{0}(\mathbf{r}-\mathbf{R}_{1})+V_{0}(\mathbf{r}-\mathbf{R}%
_{2}),  \label{Eq.:2cpot}
\end{equation}%
where $\mathbf{R}_{i}\ (i=1,2)$ denote the coordinates of the centers $%
C_{i}\ (i=1,2).$ For the ground-state wave function, we employ a linear
combination of atomic orbitals (LCAO):
\begin{equation}
\Psi (\mathbf{r})=c_{1}\Psi _{0}(\mathbf{r}-\mathbf{R}_{1})+c_{2}\Psi _{0}(%
\mathbf{r}-\mathbf{R}_{2}).  \label{Eq.:lcao}
\end{equation}%
Specifically, we will use the zero-range potential
\begin{equation}
V(\mathbf{r})=\frac{2\pi }{\kappa }\delta (\mathbf{r})\frac{\partial }{%
\partial r}r,  \label{Eq.:zrp}
\end{equation}%
whose single bound state is described by the wave function
\begin{equation}
\Psi _{0}(\mathbf{r})=\left( \frac{\kappa }{2\pi }\right) ^{1/2}\frac{1}{r}%
e^{-\kappa r},
\end{equation}%
with $\kappa =\sqrt{2|E_{0}|}$. The regularization operator $\partial
/\partial r\;r$ acts on the wave function to its right in order to satisfy
the proper boundary conditions at the origin \cite{Fermi1936}. For direct
ionization by a monochromatic linearly polarized laser field
\begin{equation}
\mathbf{A}(t)=A_{0}\cos \omega t\;\mathbf{e},  \label{Eq.:field}
\end{equation}%
the evaluation of the amplitude (\ref{Eq.:KFR}) is straightforward. Taking $%
\mathbf{R}_{1}=\mathbf{R}/2$ and $\mathbf{R}_{2}=-\mathbf{R}/2$, so that $%
\mathbf{R}$ is the internuclear distance of the two centers, one obtains by
expanding the exponent in the Volkov wave function in Eq.~(\ref{Eq.:Volkov})
into Bessel functions
\begin{eqnarray}
M_{\mathbf{p}}^{0} &=&\frac{2F}{(2\pi )^{\frac{3}{2}}}\cos \left( \frac{%
\mathbf{p}\cdot \mathbf{R}}{2}\right) \sum_{N,l}\delta \left( \frac{p^{2}}{2}%
+U_{p}+\left\vert E_{0}\right\vert -N\omega \right)  \nonumber \\
&&\times J_{l}\left( \frac{U_{p}}{2\omega }\right) J_{-(2l+N)}\left( \frac{2%
\mathbf{p\cdot e}}{\omega }\sqrt{U_{p}}\right) ,  \label{Eq.:bessel}
\end{eqnarray}%
where $U_{p}=A_{0}^{2}/4$ denotes the ponderomotive energy of the laser
field (\ref{Eq.:field}). The prefactor, which is proportional to
\begin{equation}
F=-\kappa +\exp (-\kappa R)/R,  \label{Eq.:pre-factor}
\end{equation}%
is of no relevance, since we do not attempt to calculate total ionization
rates. It is, however, worth mentioning that the limit of $R\rightarrow 0$
is not straightforward. Below we will not face this limit. For a more
detailed discussion, see, e.g., Refs.~\cite{KMJ91,KBK98}.

\begin{figure}[t]
\includegraphics[width=.4\textwidth,height=.4%
\textheight,angle=270]{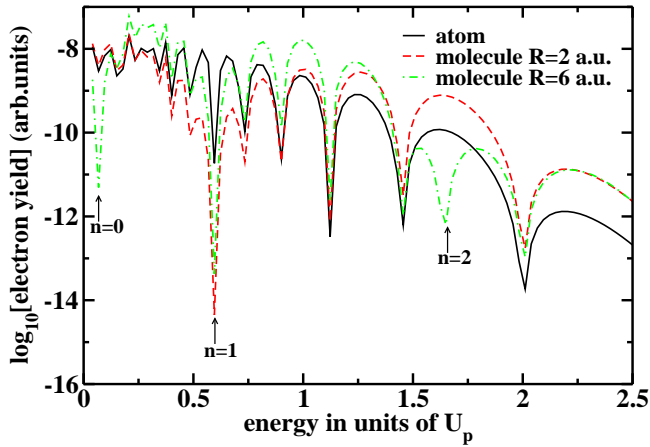}
\caption{(Color online) ATI spectra of the direct electrons in the laser polarization
direction in the atomic and molecular case, for a molecule aligned parallel
to a laser field of frequency $\protect\omega =0.058$ a.u. and ponderomotive
potential $U_{p}=2.08$ a.u. We consider a symmetric combination of atomic
orbitals [$c_{1}=c_{2}=1/\protect\sqrt{2}$ in Eq.~(5)], with ionization
potential $|E_{0}|=0.9$ a.u. In order to facilitate the comparison, the same
ionization potential $V_{0}(\mathbf{r})$ was chosen in the atomic and
molecular cases. The arrows mark the various destructive interference
energies $(n=0,1,2)$ for $R=6$ a.u..}
\label{fig.:direct}
\end{figure}
The only difference between the matrix element (\ref{Eq.:bessel}) for a
molecule and the corresponding matrix element for an atom, besides the $R$%
-dependent prefactor, is the presence of the term $\cos (\mathbf{p\cdot R}/2)
$. This term describes the interference of electron orbits with momentum $%
\mathbf{p}$ originating from one or the other center of the two-center
potential (\ref{Eq.:2cpot}). The cosine term comes from assuming a symmetric
combination of orbitals in the ground-state wave function (\ref{Eq.:lcao}),
so that $c_{1}=c_{2}=1/\sqrt{2}$. For an antisymmetric combination, so that $%
c_{1}=-c_{2}$, the cosine is replaced by a sine, leading to suppression of
electrons with low momenta due to destructive interference \cite%
{MBBF00,interfexp,doubleslit}. The interference factor $\cos (\mathbf{p\cdot
R}/2)=\cos (pR\cos \theta /2)$ yields destructive interference for electrons
with energies
\begin{equation}
E\equiv \frac{p^{2}}{2}=\frac{1}{2}\left( \frac{(2n+1)\pi }{R\cos \theta }%
\right) ^{2}  \label{Eq.:interf}
\end{equation}%
for integer $n$. An illustration of the interference effect is given in Fig.~%
\ref{fig.:direct}, which shows the spectrum of the direct electrons for an
atom and for a symmetric diatomic molecule aligned parallel to the
laser-field polarization. The clearly visible sharp dips in the spectrum,
due to destructive interference, are indicated by the arrows in the figure.
Next, we turn to the evaluation of the matrix element (\ref{Eq.:rescatter}),
which allows rescattering. With the two-center potential (\ref{Eq.:2cpot})
and the symmetric ground-state wave function (\ref{Eq.:lcao}), the matrix
element reads
\begin{widetext}
\begin{eqnarray}
M_\mathbf{p}&=&\frac{-i}{(2\pi)^{3/2}}\int_{-\infty}^{\infty}dt
\int_{-\infty}^{t}dt^{\prime}\int
d^{3}r\int d^{3}r^{\prime}\; e^{-i\mathbf{p}\cdot\mathbf{r}}
e^{\frac{i}{2}\int_{-\infty}^{t}d\tau(\mathbf{p}+\mathbf{A}(\tau))^{2}} 
\left[V_{0}(\mathbf{r}+\mathbf{R/2})+V_{0}(\mathbf{r}-\mathbf{R/2})\right]
\nonumber \\
&\times& U^{(V)}(\mathbf{r}t;\mathbf{r^{\prime}}t^{\prime})
\left[V_{0}(\mathbf{r^{\prime}}+\mathbf{R/2})
+V_{0}(\mathbf{r^{\prime}}-\mathbf{R/2})\right]
\left[\Psi_{0}(\mathbf{r^{\prime}}+\mathbf{R/2}) +
\Psi_{0}(\mathbf{r^{\prime}}-\mathbf{R/2})\right] e^{i|E_{0}|t'}\;.
\label{M}
\end{eqnarray}
\end{widetext}For the zero-range potential (\ref{Eq.:zrp}), the integrations
over space can be carried out easily \cite{LKKB97}, which leaves a
two-dimensional integral over the ionization time $t^{\prime }$ and the
rescattering time $t$. For finite-range potentials, one may proceed by
introducing form factors and employing saddle-point methods. In this case,
the single-center pre\-factors cause an overall decrease of the yield for
increasing photoelectron energy. There are, however, no significant
changes in the interference patterns in comparison with the zero-range case,
since these prefactors do not influence the action or the cosine factor. For
a detailed discussion of the single-atom case, see, e.g., Ref.~\cite%
{Faria2002}.

We split the eight integrals into those where the electron rescatters off
the same center from which it was ionized $(\mathbf{r}=\mathbf{r^{\prime }}%
=\pm \mathbf{R/2})$ and those where it rescatters off the opposite center $(%
\mathbf{r}=-\mathbf{r^{\prime }}=\pm \mathbf{R/2})$. We refer to the
respective terms by $M^{ij}$ where $i=+,-$ denote the center of ionization
and $j=+,-$ the center of rescattering. The integrals $M^{++}$ and $M^{--}$,
which specify the electrons coming from and rescattering off the same
center, are essentially identical to the corresponding results for an atom
\cite{LKKB97}. The structure of the molecule is reflected in the integrals $%
M^{+-}$ and $M^{-+}$, which characterize the electrons that experience the
presence of both centers. Evaluating the remaining two integrals over $t$
and $t^{\prime }$, we substitute $t^{\prime }=t-\tau $. The doubly infinite
integral over $t$ then yields a $\delta $-function expressing energy
conservation, while the semi-infinite integration over $\tau $ has to be
calculated numerically. Expanding all oscillating exponents in terms of
Bessel functions, we obtain for the transition amplitudes

\begin{widetext}
\begin{eqnarray}
M_\mathbf{p}^{++} + M_\mathbf{p}^{--} & = & 2F \cos\left(\frac{\mathbf{p\cdot R}}{2}\right)
\sum_{N}\delta\left(\frac{p^{2}}{2}+U_{p}+|E_{0}|-N\omega\right)
\sum_{l=-\infty}^{\infty}J_{-(2l+N)}\left(\frac{2\mathbf{p \cdot e}}{\omega}\sqrt{U_{p}}\right)
\nonumber\\ &&\times \int_{0}^{\infty}d\tau \left(\frac{i}{2\pi\tau}\right)^{3/2} \left\{
 e^{-i(|E_{0}|\tau+l\alpha)}e^{-iU_{p}\tau[1-(\frac{\sin\omega\tau/2}{\omega\tau/2})^{2}]}
 J_{l}\left(\frac{U_{p}}{2\omega}A\right)
         -J_{l}\left(\frac{U_{p}}{2\omega}\right) \right\}\nonumber\\
&=& 2F \cos\left(\frac{\mathbf{p\cdot R}}{2}\right) M_\mathbf{p}^{(\mathrm{atom})},  \label{m++--}
\end{eqnarray}
\begin{eqnarray}
M_\mathbf{p}^{+-} & =& F e^{-i\frac{\mathbf{p\cdot R}}{2}}
\sum_{N,l}\delta\left(\frac{p^{2}}{2}+U_{p}+|E_{0}|-N\omega\right)\int_{0}^{\infty}d\tau
\left(\frac{i}{2\pi\tau}\right)^{3/2}\nonumber\\ & &
\times \left\{ e^{i\frac{R^{2}}{2\tau}}
 e^{-i[|E_{0}|\tau+l\alpha+(2l+N)\beta_-]}
 e^{-iU_{p}\tau[1-(\frac{\sin\omega\tau/2}{\omega\tau/2})^{2}]}
 J_{l}\left(A\frac{U_{p}}{2\omega}\right)J_{-(2l+N)}
 \left(\frac{2\sqrt{U_{p}}}{\omega}B_-\right)
    \right\},  \label{m+-}
\end{eqnarray}
\begin{eqnarray}
M_\mathbf{p}^{-+} & =& F e^{i\frac{\mathbf{p\cdot R}}{2}}
\sum_{N,l}\delta\left(\frac{p^{2}}{2}+U_{p}+|E_{0}|-N\omega\right)\int_{0}^{\infty}d\tau
\left(\frac{i}{2\pi\tau}\right)^{3/2}\nonumber\\ & &
\times\left\{e^{i\frac{R^{2}}{2\tau}}
e^{-i[|E_{0}|\tau+l\alpha+(2l+N)\beta_+]}
e^{-iU_{p}\tau[1-(\frac{\sin\omega\tau/2}{\omega\tau/2})^{2}]}
 J_{l}\left(A\frac{U_{p}}{2\omega}\right) J_{-(2l+N)}
 \left(\frac{2\sqrt{U_{p}}}{\omega}B_+\right)
\right\}. \label{m-+}
\end{eqnarray}
\end{widetext}
The real quantities $A$, $B_\pm$ and the phases $\alpha$ and $\beta_\pm$ are
defined by
\begin{eqnarray}
Ae^{-i\alpha} &=& e^{-2i\omega \tau} +\frac{8i}{\omega\tau} \sin^2\frac{%
\omega\tau}{2} e^{-i\omega\tau},  \label{defA} \\
B_\pm e^{-i\beta_\pm}&=& \mathbf{p \cdot e} \pm \frac{\mathbf{R \cdot e}}{%
\tau}[i\sin \omega \tau -(1-\cos \omega \tau)].  \label{defB}
\end{eqnarray}
Upon $\mathbf{R}\rightarrow -\mathbf{R}$, we have $B_{\pm }\exp (-i\beta
_{\pm })\rightarrow B_{\mp }\exp (-i\beta _{\mp })$. Consequently, the
matrix element $M_{\mathbf{p}}$ does not change when $\mathbf{R}\rightarrow -%
\mathbf{R}$. The complete matrix element is the sum of the terms (\ref{m++--}%
) -- (\ref{m-+}),
\begin{equation}
M_{\mathbf{p}}=M_{\mathbf{p}}^{++}+M_{\mathbf{p}}^{--}+M_{\mathbf{p}%
}^{+-}+M_{\mathbf{p}}^{-+}.  \label{mtotal}
\end{equation}%
The first two terms describe electrons originating from and
rescattering off the same center. They are proportional to the atomic ionization amplitude $M_\mathbf{p}^{(\mathrm{atom})}$ \cite{LKKB97} multiplied by the wave-function overlap $F$  and the two-center interference factor $\cos (\mathbf{p\cdot R}/2)$, which we observed for the
direct electrons in Eq.~(\ref{Eq.:bessel}). The behavior of the exchange
terms $M_{\mathbf{p}}^{+-}$ and $M_{\mathbf{p}}^{-+}$ is more complicated
and will be discussed below.

The transition amplitude $M_\mathbf{p}$ simplifies enormously when the
molecule is aligned perpendicularly to the field so that $\mathbf{R}\cdot%
\mathbf{e}=0$. Equation (\ref{defB}) shows that in this case $B_+=B_-$ and $%
\beta_+=\beta_-=0$. Hence, the integrals on the right-hand side of Eqs.~(\ref%
{m+-}) and (\ref{m-+}) are equal and $M_\mathbf{p}^{+-}+M_\mathbf{p}^{-+}$
becomes proportional to $2\cos(\mathbf{p}\cdot\mathbf{R}/2)$ just like $M_%
\mathbf{p}^{++}+M_\mathbf{p}^{--}$. If, in addition, the electron is emitted
perpendicularly to the field so that also $\mathbf{p}\cdot\mathbf{e}=0$,
then $B_+=B_-=0$ and we have $M_\mathbf{p}=0$ unless $N$ is even.

Substituting $\tau \to \tau/\omega$ in Eqs.~(\ref{m++--})--(\ref{defB}) one
can see that the amplitudes $M^{ij}_\mathbf{p}$ and their sum $M_\mathbf{p}$
depend on the parameters of the problem through the dimensionless quantities
$p^2/\omega$, $U_p/\omega$, and $\omega R^2$ when the relative orientations
of the vectors $\mathbf{R}$, $\mathbf{p}$, and $\mathbf{e}$ are kept fixed.

\section{Photoelectron spectra}

\label{results}

In this section we discuss the ATI spectra computed employing the transition
matrix elements (\ref{Eq.:KFR}) and (\ref{Eq.:rescatter}), and a symmetric
combination of equivalent centers [$c_{1}=c_{2}=1/\sqrt{2}$ in Eq.~(\ref%
{Eq.:lcao})]. For the sake of simplicity, in the comparison between the
atomic and molecular case the same ionization potential $V_{0}$ is chosen.
Specifically, we take $|E_{0}|=0.9$ a.u. in Eqs. (5) and (7) throughout \cite%
{footnIp}. In Sec.~\ref{alignment}, we perform a detailed analysis of the
interference patterns with respect to the molecular orientation,
rescattering scenarios, and the direction of electron emission, while in
Sec.~\ref{contourplots} we provide a recipe for measuring the internuclear
distance from an analysis of the interference patterns in the angle-resolved
spectra.

\subsection{Analysis of the interference patterns}

\label{alignment}

\begin{figure}[tbp]
\includegraphics[width=.4\textwidth,height=.4%
\textheight,angle=270]{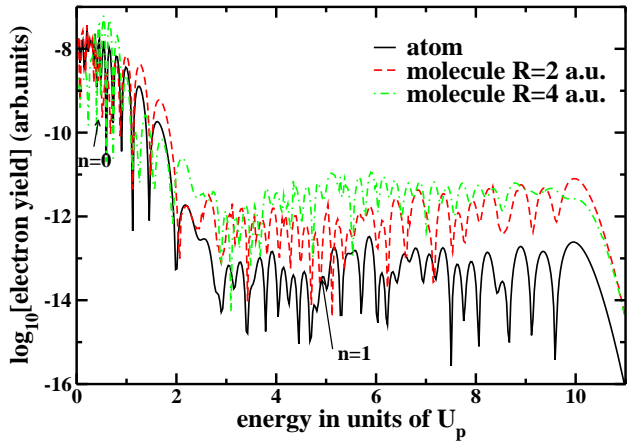}
\caption{(Color online) Comparison of the complete ATI spectra consisting of direct and
rescattered electrons in the atomic and the molecular case for internuclear
distances of $R=2$ a.u. and $R=4$ a.u.. The molecule is aligned parallel to
the laser-polarization direction, and the electrons are emitted in the same
direction. The arrows mark the destructive interferences $(n=0,1)$ of the
molecule for $R=2$ a.u.. The destructive interference for $n=1$ is already in the
plateau region, where the role of exchange terms becomes important. The
remaining parameters are the same as in Fig. 1. }
\label{fig.:direct+rescatter}
\end{figure}

\begin{figure}[tbp]
\includegraphics[width=.4\textwidth,height=.4%
\textheight,angle=270]{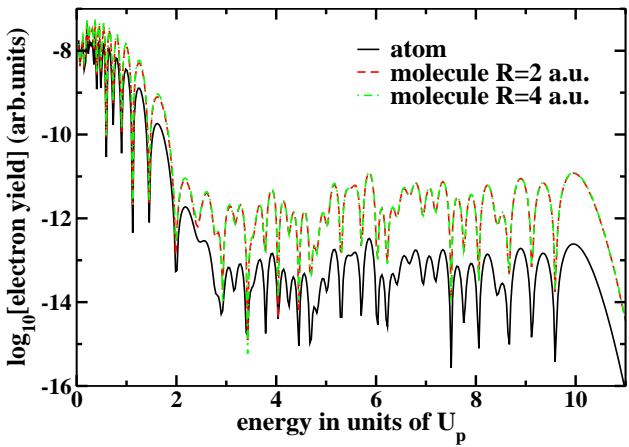}
\caption{ (Color online) The same as Fig.~\protect\ref{fig.:direct+rescatter} but with the
molecular axis perpendicular to the laser polarization. The electrons are
emitted parallel to the laser polarization.}
\label{fig.:orientation}
\end{figure}

As a first step, we investigate how the interference patterns are influenced
by the orientation of the molecule with respect to the laser-polarization
direction. Such results are displayed in Figs. \ref{fig.:direct+rescatter}
and \ref{fig.:orientation} for parallel and perpendicular orientations,
respectively. In both cases, we compare the entire ATI spectrum consisting
of the direct and the rescattered electrons in the atomic and the molecular
case. Unless stated otherwise (cf. Fig. \ref{fig.:emission_angle}), we
consider electron emission in the laser-polarization direction.

As expected from Eq.~(\ref{Eq.:bessel}), for energies smaller than $2U_{p},$
the main contributions to the yield come from the direct-ionization matrix
element (\ref{Eq.:bessel}). Apart from the interference-related factor of $%
\cos(\mathbf{p}\cdot\mathbf{R}/2)$, the transition matrix element is
identical to that obtained for a single atom (cf. Fig. 1). This factor is
responsible for the sharp interference dips at the positions given by Eq.~(%
\ref{Eq.:interf}). In the plateau energy region, however, the spectra depend
on the laser-field polarization in a more complex way, as will be discussed
next. In case the molecule is aligned parallel to the laser-field
polarization (Fig. \ref{fig.:direct+rescatter}), the plateau is strongly
enhanced in the molecular case, and the structure of the spectrum is very
different from the atomic case and dependent on the internuclear distance $R$%
. Indeed, inspection of the exchange integrals (\ref{m+-}) and (\ref{m-+})
does not reveal any simple dependence on the internuclear distance.
Generally, for the molecular case there are more pathways into a given final
state. For our case of a two-center potential, there are four pathways 
in place of one for the atomic case. 
If they add coherently, a significant enhancement can result, 
ideally by a factor of 16, which is roughly what is observed in Fig.~
\ref{fig.:direct+rescatter} before the cutoff.  
The structure caused by the cosine factor is
suppressed in the plateau region. This is caused by the contribution of the
processes in which the electron is ejected from one center and rescatters
off the other. Such processes correspond to the transition amplitudes $%
M_{p}^{-+}$ and $M_{p}^{+-}$, which do not exhibit the proportionality to
the cosine that is characteristic of the the one-center scattering
amplitudes $M_{p}^{++}$ and $M_{p}^{--}$. A further particular feature
observed in this case is the displacement of the cut-off to higher energies
with increasing internuclear distance. This can be understood by the fact
that an electron that moves from one center to the other may gain more
energy from the field since it may be accelerated over a longer distance
before it recollides.

A strikingly different behavior is observed if the molecule is aligned
perpendicular to the direction of the laser field  so that $\mathbf{R}\cdot\mathbf{e}=0$.  In Fig.~\ref%
{fig.:orientation} we consider the case where the electron is emitted in the direction of the laser polarization so that $\mathbf{p}\cdot\mathbf{R}=0$, too. In this case, there is a general enhancement of the ATI
yield in comparison to that of a single atom by roughly a factor of two for
the direct electrons and a much larger factor for the rescattered electrons.
Notice that the molecular spectrum is practically independent of the internuclear
distance \cite{Hetzheim2005}, since the $R$ dependence of the prefactor (\ref{Eq.:pre-factor}) is weak for $R\agt 2$ and the exponential of  $R^2/(2\tau)$ is small for the values of $R$ that we consider and the values of $\tau$ that give the dominant contributions to the integral. The entire ATI
spectrum does not show any interference structure, since the contributions from the two centers add constructively for $\mathbf{p}\cdot\mathbf{R}=0$.  In fact, the cosine term in the matrix elements $M_{\mathbf{p}}^{0}$%
, $M_{p}^{++}$ and $M_{p}^{--}$ simply reduces to one and the spectrum
therefore looks like that of an atom. Specifically within the plateau, by
symmetry the contributions from the two centers always interfere
constructively. This results in a spectrum that is largely independent on
the internuclear separation $R$, except that the plateau is enhanced,
compared with the atomic case, by the existence of four pathways. Formally,
this can be understood as discussed above at the end of Sec. II.

\begin{figure}[tbp]
\includegraphics[width=.4\textwidth,height=.4%
\textheight,angle=270]{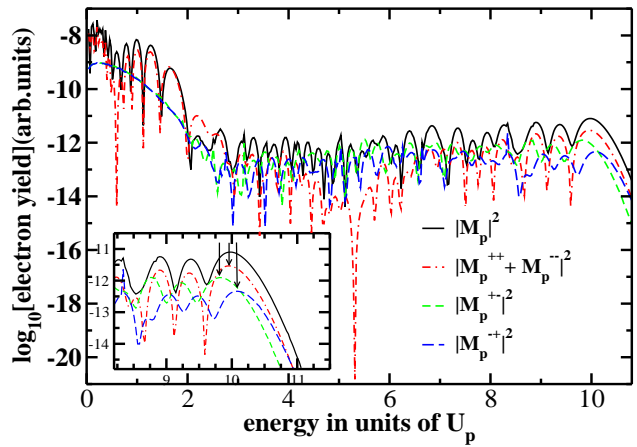}
\caption{(Color online) Individual contributions of the various rescattering scenarios to the total amplitude (\protect\ref%
{mtotal}) for a diatomic molecule with internuclear distance $R=2$ a.u.  
aligned parallel to the laser-field polarization, for the same molecular and
field parameters as in Fig.~\ref{fig.:direct+rescatter}. The electrons are emitted in the polarization direction. The arrows mark the respective cutoff
energies for the various transition amplitude matrix elements. The inset at the lower left is an enlargement of the region near the cutoff where the direct terms and the exchange terms differ in a characteristic fashion, allowing for the determination of the internuclear separation. }
\label{fig.:contribution_M}
\end{figure}

\begin{figure}[tbp]
\includegraphics[width=0.4\textwidth,height=.4%
\textheight,angle=-90]{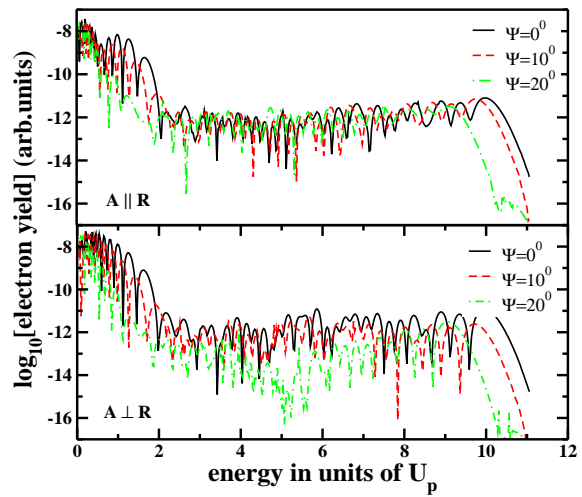}
\caption{(Color online) Electron yield for the parameters of Fig. 1 and internuclear
distance $R=2$ a.u. for different emission angles $\protect\psi$ with respect to
the polarization of the laser field. The molecule is aligned parallel
(perpendicular) to the laser field in the upper (lower) panel.}
\label{fig.:emission_angle}
\end{figure}

For arbitrary $\mathbf{p}\cdot\mathbf{R}$, if the electron is emitted
perpendicular to the laser polarization so that $\mathbf{p\cdot e}=0$, then
we see from Eqs.~(\ref{defA}) and (\ref{defB}) that $B_{+}=B_{-}\!$, while $%
\beta _{-}=\beta _{+}+\pi $. The sum of the two exchange terms then goes
like $\cos (\mathbf{p\cdot R}/2)$ for even $N$ and like $\sin (\mathbf{%
p\cdot R}/2)$ for odd $N$. This holds regardless of the orientation of the
molecule. If the laser polarization is perpendicular to both the electron
momentum and the internuclear axis, then $B_{+}=B_{-}=0$. This implies that $%
M_{\mathbf{p}}$ is nonzero only for integer $N$. Each other electron peak is
missing.
Next, we discuss the interference pattern in more detail by analyzing the
individual contributions to the transition matrix element. In Fig.~\ref%
{fig.:contribution_M}, we separately investigate the individual
contributions to the amplitude (\ref{mtotal}). If only $M_{p}^{++}$ and $M_{p}^{--}$ are taken, a very
pronounced mininum is observed near $5U_{p}.$ These matrix elements
correspond to the case in which the electron is ejected from and rescatters
off the same center, so that the minimum is due to the term $\cos (\mathbf{%
p\cdot R}/2)$. In the full spectrum $|M_{\mathbf{p}}|^{2},$ however, this
minimum is absent because it is filled by the contributions from the
exchange terms $M_{p}^{-+}$ and $M_{p}^{+-}$. For a given orientation of the
molecule with respect to the laser field and for fixed momentum $\mathbf{p,}$
Fig.~\ref{fig.:contribution_M} shows that the contribution $|M_{p}^{-+}|^{2}$
of the scenario in which the electron is freed at the center $C_{1}$ and
rescatters off at the center $C_{2}$ is different from that of $%
|M_{p}^{+-}|^{2}$ where it is released at $C_{2}$ and rescatters at $C_{1}$.
The same has been observed for high-order harmonic generation in a
two-center system \cite{KBK98}.

The imprints of interference can still be observed if the electron is
emitted away from the laser polarization direction. This will cause,
however, an overall decrease in the photoelectron energies for both the
direct and the rescattered electrons. Examples are presented in Fig.~\ref%
{fig.:emission_angle}. This behavior is known from atomic ionization, and
its origin is the same in the molecular case; for a discussion, see, e.g., 
Ref.~\cite{LKKB97}.

\subsection{ Determining the internuclear distance}

\label{contourplots}

\begin{figure}[h]
\includegraphics[width=0.4\textwidth,height=.2\textheight]{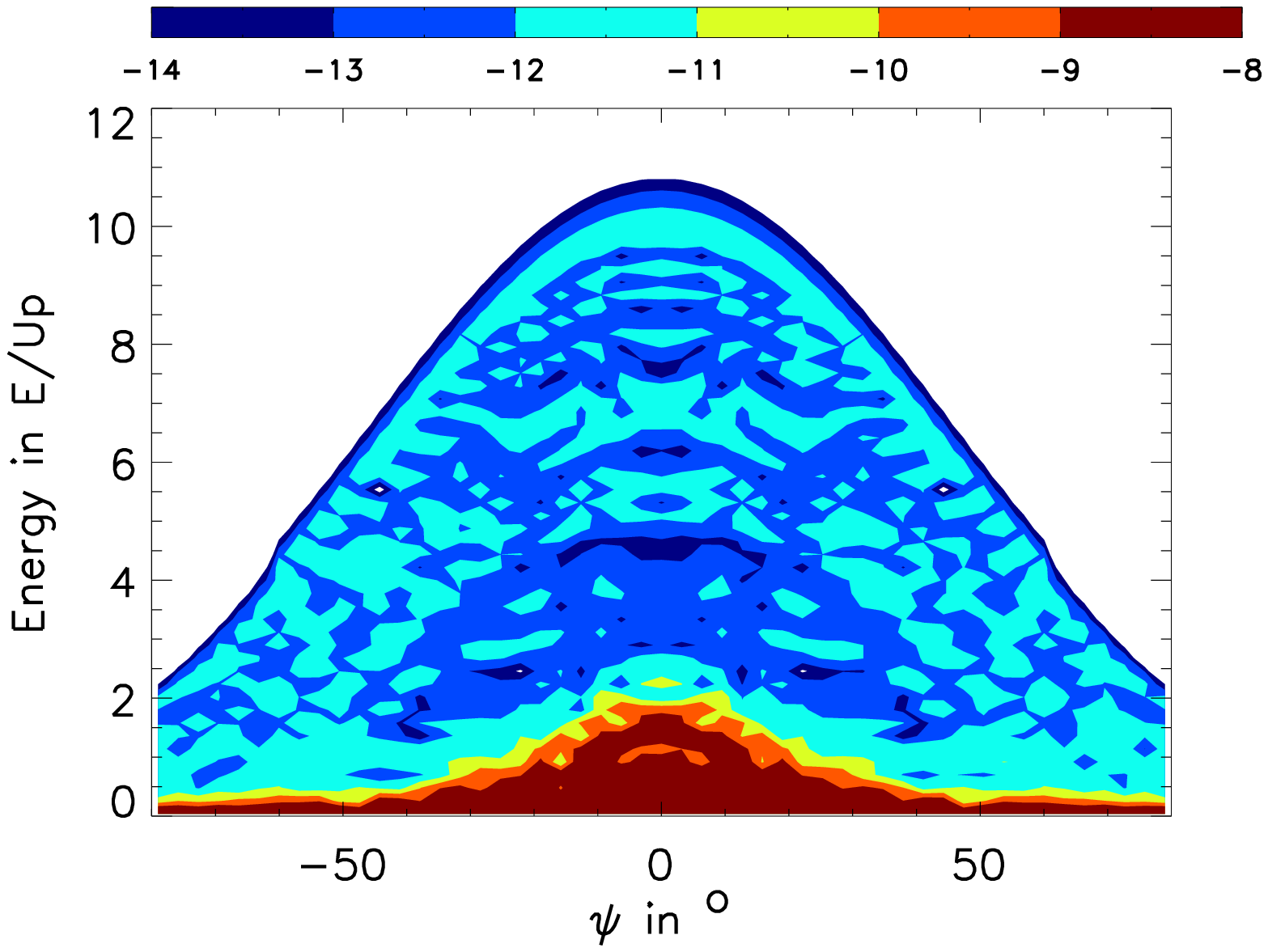}
\vspace{\floatsep} \includegraphics[width=0.4\textwidth,height=.2%
\textheight]{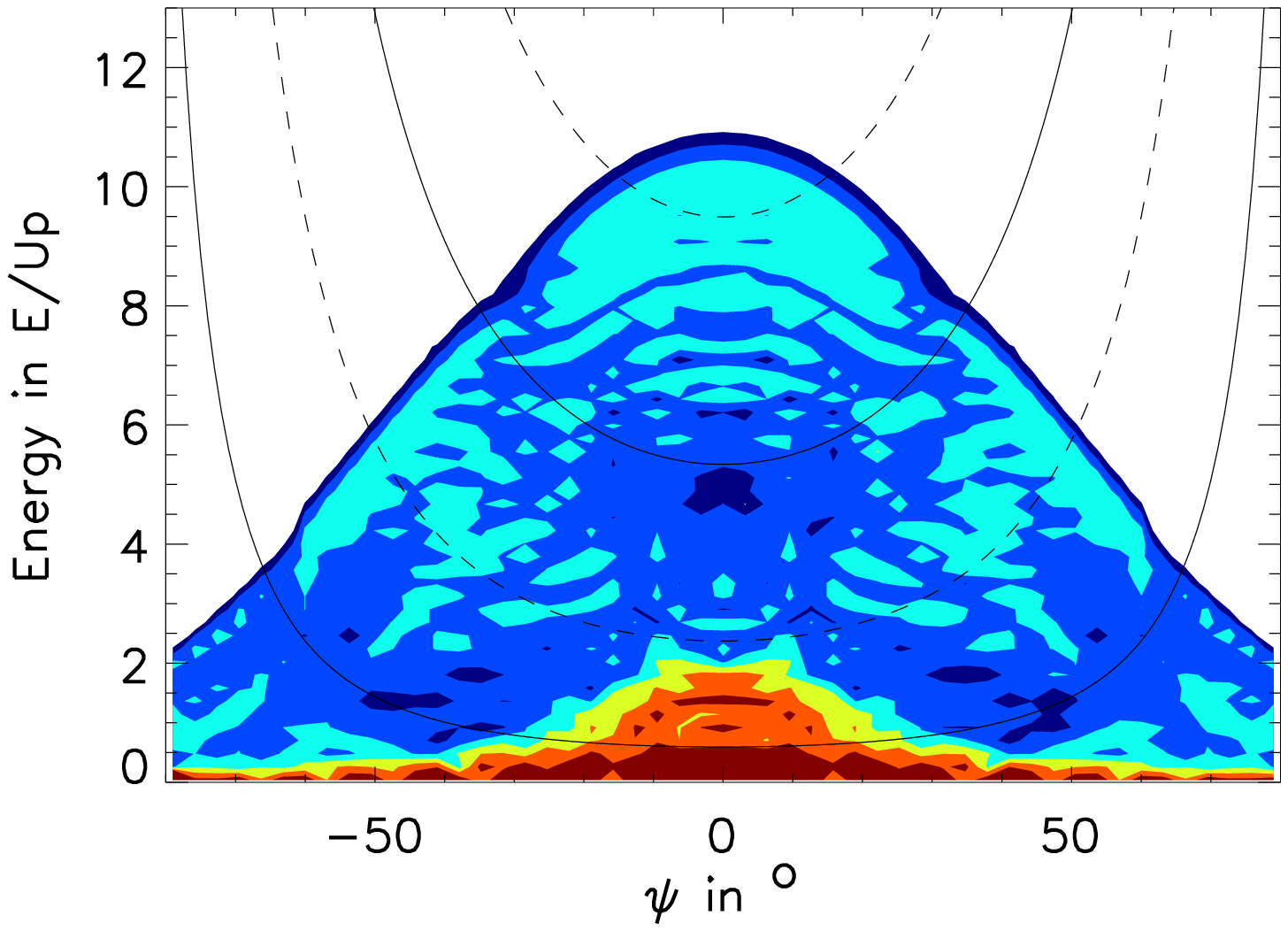} \vspace{\floatsep} %
\includegraphics[width=0.4\textwidth,height=.2%
\textheight]{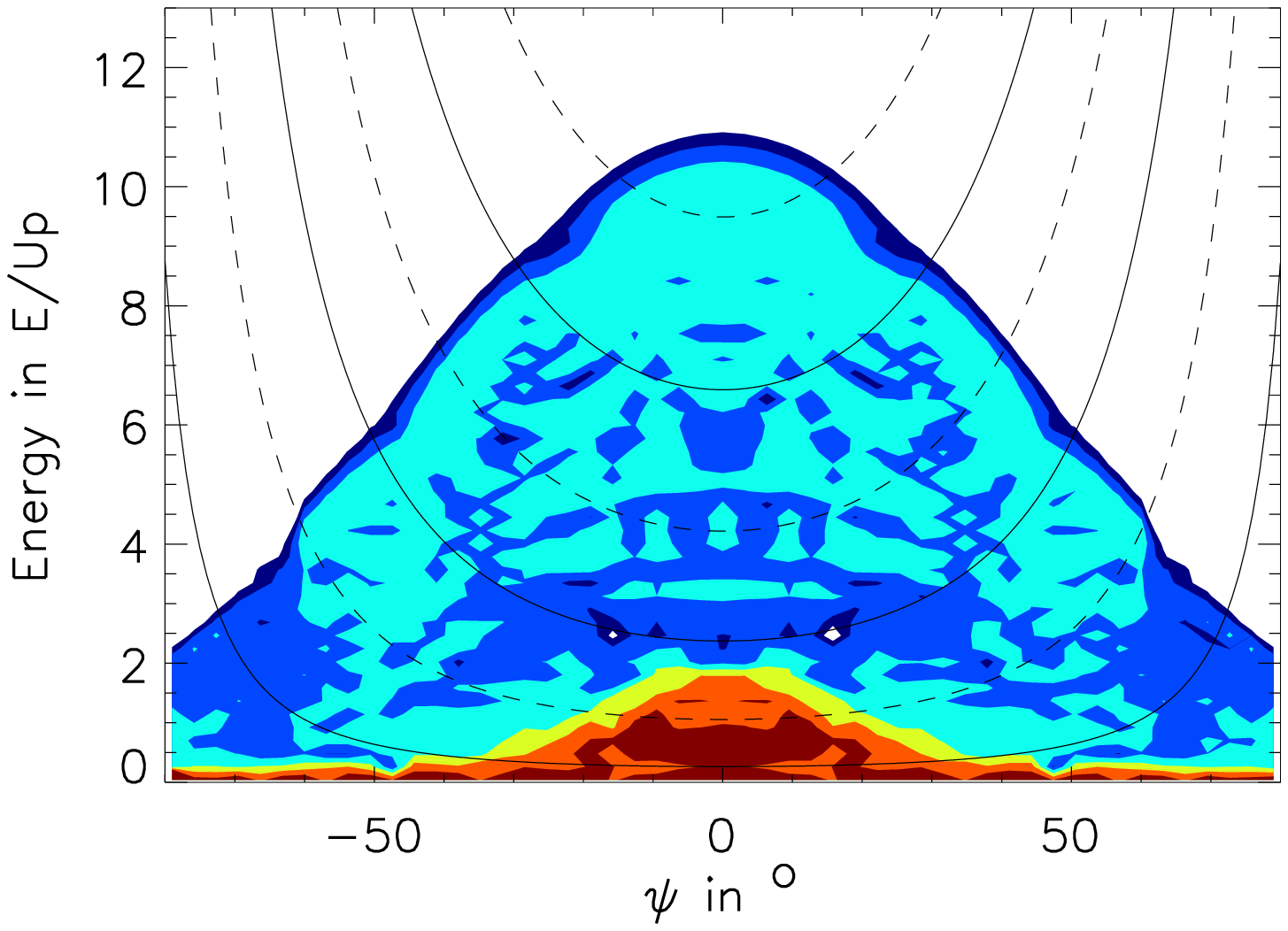}
\caption{{(Color online)\protect\small { Angle-resolved ATI spectra on a logarithmic scale
for a diatomic molecule with internuclear distances $R=2$ a.u. and $R=3$
a.u. (middle and bottom panels, respectively), aligned parallel to the
laser-field polarization, compared to the single-atom case (upper panel).
The binding energy is $E_0=0.9$ a.u. in all cases, and the laser frequency
and the ponderomotive potential are $\protect\omega=0.058$ a.u. and $%
U_{p}=2.08$ a.u., respectively. The plotted lines depict the minima (solid
lines) and the maxima (dashed lines) of the energy distribution given by Eq.
(\protect\ref{Eq.:interf}).}}}
\label{fig.:angle-resolved}
\end{figure}
\newpage

\qquad \qquad \qquad \qquad \qquad \qquad \qquad \qquad \vspace*{0.6cm}
\begin{figure}[h]
\includegraphics[width=0.4\textwidth,height=.2%
\textheight]{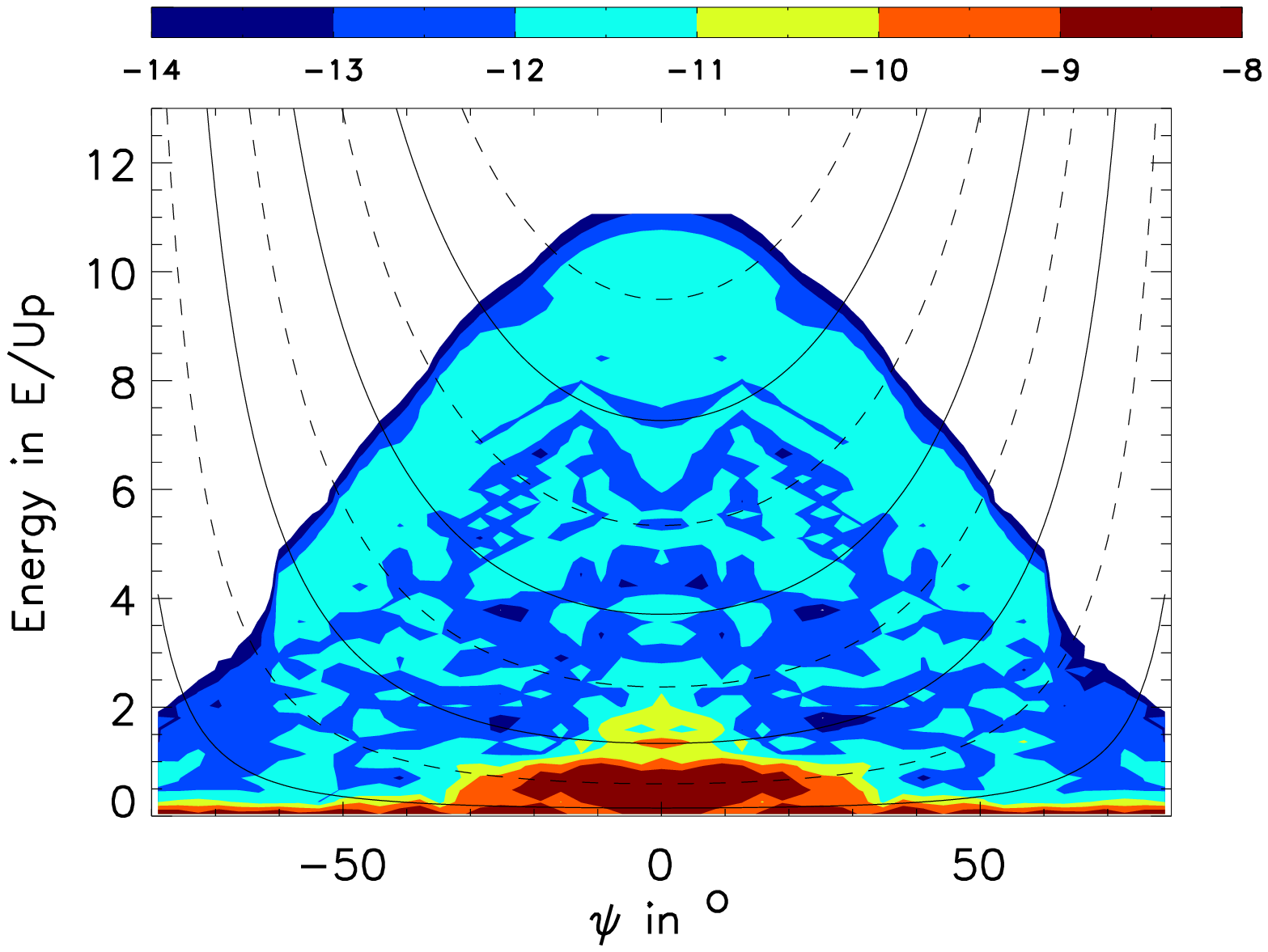} \vspace{\floatsep} %
\includegraphics[width=0.4\textwidth,height=.2%
\textheight]{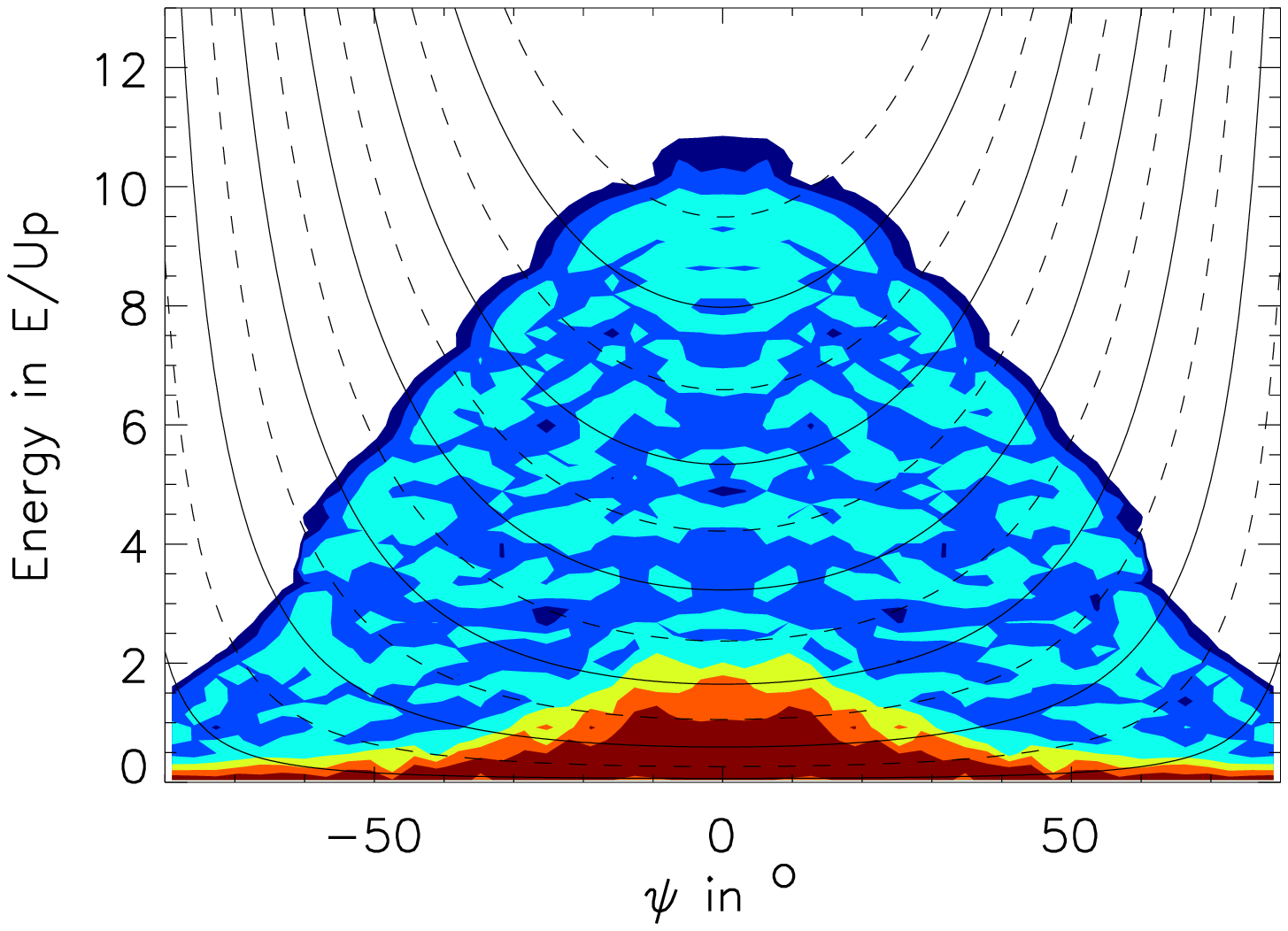} \vspace{\floatsep} %
\includegraphics[width=0.4\textwidth,height=.2%
\textheight]{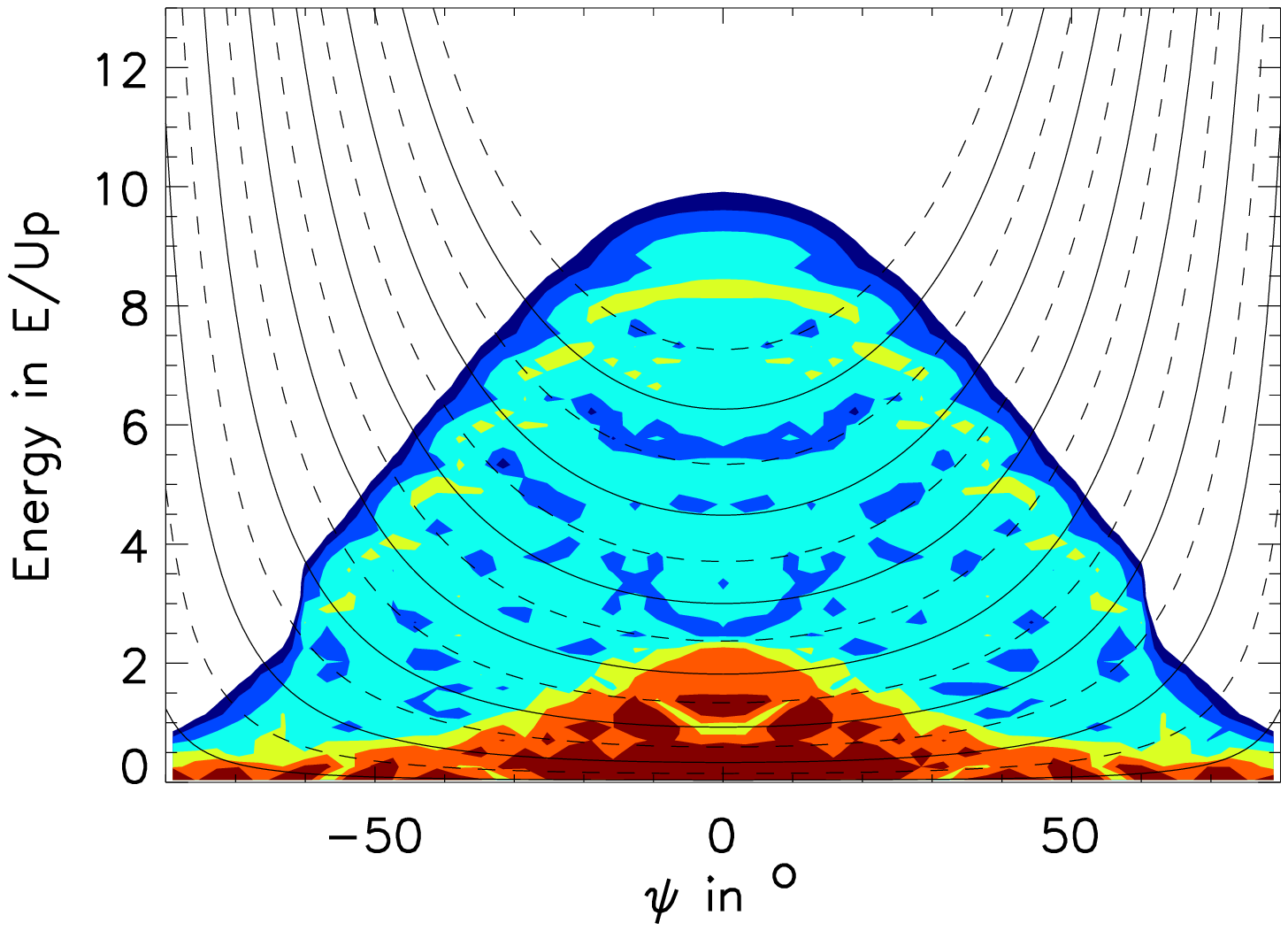}
\caption{(Color online){\protect\small { Angle-resolved ATI spectra on a logarithmic scale
for a diatomic molecule with ionization potential $E_{0}=0.9$ a.u. and
internuclear distances $R=4$ a.u., $R=6$ a.u. and $R=8$ a.u. (upper, middle
and bottom panels, respectively), aligned parallel to the laser-field
polarization. The field parameters are the same as in the previous figure.
The plotted lines depict the minima (solid lines) and the maxima (dashed
lines) of the energy distribution given by Eq. (\protect\ref{Eq.:interf}).}}}
\label{fig.:angle-resolved1}
\end{figure}

\newpage \bigskip\

For a complete picture of the angle-resolved ATI spectrum, not restricted to
emission in particular directions, we will now present density plots. While
they invariably imply loss of fine details and depend on the positioning and
gradient of the false-color scale, they give a comprehensive overview of the
general structure. We restrict ourselves to the case of parallel alignment
\footnote{%
If one changes the orientation of the molecule, the situation becomes
different. For small angles the electrons with the maximal energy are still
observable in the direction of the laser field, but there exists a further
maximum of electrons with a certain momentum in the opposite direction as a
result of the alignment of the molecule \cite{Hetzheim2005}. With increasing
angle of the molecular axis with respect to the laser polarization direction,
this local maximum will move over the entire angle-resolved ATI spectrum.}.
In this case the spectrum is symmetrical with respect to the internuclear
axis. It is obvious that the electrons with maximal kinetic energy will be
detected in the direction of the laser field.

\begin{figure}
\includegraphics[width=0.4\textwidth]{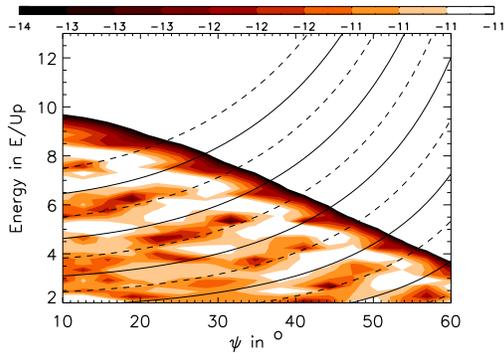}
\caption{(Color online) Enlargement of a limited energy-angle region of Fig.~\ref{fig.:angle-resolved1} for $R=8$ a.u. (lowest panel) with increased resolution. The indents of Fig.~\ref{fig.:angle-resolved1} are distinctly visible as valleys deeply cut into the high ridge that precedes the cutoff.}
\label{fig8}
\end{figure}

The angle-resolved spectra displayed in Figs.~\ref{fig.:angle-resolved} and %
\ref{fig.:angle-resolved1} are very intricate and do not exhibit any simple
structures. They depend strongly on the internuclear separation but do not,
on a first inspection, lend themselves in any obvious way to the assignment
of a specific value of $R$ to a given spectrum. Especially, owing to the
presence and magnitude of the exchange terms (\ref{m+-}) and (\ref{m-+}),
the two-center interference, which is expressed in the $\cos(\mathbf{p}\cdot%
\mathbf{R}/2)$ term, is not immediately visible. However, looking more
closely, one can observe a very distinct manifestation of this term just
near the classical boundary of the spectrum. Roughly, the latter agrees with
the boundaries of the colored areas in the various panels of Figs.~\ref%
{fig.:angle-resolved} and \ref{fig.:angle-resolved1}. We observe
well-defined indents in the overall smooth curve that defines the classical
boundary. The positions of these indents and, especially, their separations
agree quite well with the interference minima predicted by Eq.~(\ref%
{Eq.:interf}). The figures show that the separation $\delta E$ between the
indents (on the scale of $U_p$) monotonically decreases with increasing $R$.
Hence, by comparing a measured angle-resolved spectrum with Figs.~\ref%
{fig.:angle-resolved} and \ref{fig.:angle-resolved1} we can infer the
internuclear separation. For the parameters underlying Figs.~\ref%
{fig.:angle-resolved} and \ref{fig.:angle-resolved1}, the resulting function
$R(\delta E)$ is given in Table \ref{tab:dist}. Fig.~\ref{fig8} exhibits an enlargement of the relevant area around the classical cutoff for the case of $R=8$ a.u. with a higher resolution of the electron yield. The indents are very clearly visible like valleys that cut into the drop of a plateau on a topographical map.

An analytical formula for $R(\delta E)$ can be gained from an analytical
formula for the classical cutoff energy $E(\theta)$ as a function of the
angle $\theta$. Intersecting this with the energies of the interference
minima given by Eq.~(\ref{Eq.:interf}) allows one to determine the function $%
R(\delta E)$ in dependence of the parameters of the problem. For the case of
an atom, such a formula for $E(\theta)$ is actually known \cite{haso07}. At
least for $R\le 6$ a.u., Figs.~\ref{fig.:angle-resolved} and \ref%
{fig.:angle-resolved1} show that this classical boundary does not depend
very strongly on the internuclear separation $R$, so that the atomic result
could be employed. However, even with this simplification, the resulting
formula is quite complicated and we refrain from presenting it here.

The question arises of why near the classical boundary the interference term
$\cos(\mathbf{p}\cdot\mathbf{R}/2)$ roughly multiplies the angle-resolved
spectrum, like it does for the direct electrons. The answer can be inferred from Fig.~\ref{fig.:contribution_M}. The total ionization amplitude $M_\mathbf{p}$ is the superposition (\ref{m++--}) of four different scenarios such that the electron starts from and rescatters off one or the other center. The  two contributions  (\ref{m++--}) where they start from and rescatter off the same center are identical except for the geometrical phase, which leads to the cosine in Eq.~(\ref{m++--}).  In contrast, the other two contributions (\ref{m+-}) and (\ref{m-+}) are uncorrelated since they are generated by geometrically different scenarios. Their magnitudes are different and almost nowhere do they exhibit a significant constructive interference. The two contributions (\ref{m++--}) are individually large when the long orbit and the short orbit add constructively, as is the case specifically just before the classical cutoff. In this case, they dominate the other two terms (\ref{m+-}) and (\ref{m-+}) by a factor of the order of 2 to 4. Hence the complete spectrum distinctly exhibits the geometrical interference, which is expressed in the factor $\cos(\mathbf{p}\cdot\mathbf{R}/2)$. 

\begin{table}[tbh]
\begin{tabular}{l|l}
\hline\hline
Internuclear distances & Energy differences of the indents \\
of the molecule & at the spectral boundary \\ \hline\hline
$R=2$ a.u. & $\delta E\approx 4.5 U_{p}$ \\
$R=3$ a.u. & $\delta E\approx 3.2 U_{p}$ \\
$R=4$ a.u. & $\delta E\approx 2.4 U_{p}$ \\
$R=6$ a.u. & $\delta E\approx 1.3 U_{p}$ \\
$R=8$ a.u. & $\delta E\approx 1.1 U_{p}$%
\end{tabular}%
\caption{{\protect\small {Energy differences between adjacent indents around
the classical boundary of the angle-resolved spectrum of Fig.~\protect\ref%
{fig.:angle-resolved}. The differences are taken, for each internuclear
separation, by starting with the first indent for $\Psi \geq 0^{\circ }$ as
a function of the energy.}} }
\label{tab:dist}
\end{table}

\section{Conclusions}

\label{conclusions}

We have analyzed ATI spectra for a two-center molecule in a linearly
polarized laser field. The terms of the two-center wave function
contributing to the interference structure within the SFA formalism could be
identified as well as the absence of the interference structure throughout
most of the plateau region. We have shown that the angle-resolved spectra
can be used to determine the internuclear distance of a molecule aligned
with the laser field, by reading off the energy differences between
subsequent interference minima at the classical boundary of the spectrum.

The validity of this method depends upon how close to reality is the
angle-resolved spectrum calculated for our model molecule. Certainly, the
spectrum of the \textit{direct} electrons cannot be trusted for this
purpose. However, \textit{high-order} ATI of an atom is well described by
the SFA and a zero-range potential, especially near the classical boundary
\cite{LKKB97}; for a comparison of spectra calculated from the SFA with the
solution of the time-dependent Schr\"odinger equation, see Ref. \cite{BMB06}
for the case of an atom. Experimentally, application of the method requires
a high detection efficiency that allows one to obtain a sufficient number of
counts down to the classical boundary.

The problem of how to extract the internuclear separation from a diffraction
pattern has been addressed by a different method in Ref.~\cite{Hu2005},
employing the numerical solution of the time-dependent Schrdinger equation.
In \cite{Hu2005}, however, it is necessary to compute a radial distribution
function from the diffraction intensity, whereas, with the method discussed
in this paper, one may determine the internuclear distances directly from
the spectra. The method suggested in Ref.~\cite{Hu2005} has the advantage
that it analyzes direct electrons and, therefore, does not require
exceptionally high detection efficiency.

Finally, in a real physical system, there exist additional effects,
which have not been incorporated in this model and may alter the
interference patterns. Molecular vibration, for instance, causes an
intensity loss in the high-harmonic signal \cite{SFAvibrLein}, which
may lead to a blurring in the patterns. However, recently, numerical
ATI computations in which such an effect is included have shown that
for H$_2^+$ the angle-dependent interference patterns related to 
the double-slit
physical picture remain distinguishable in the case considered 
\cite{fewcycvibr}. Generally, the amount of  blurring depends on the rigidity of the vibrational potential. Since the period of vibrations is much longer than the laser period, with a few-cycle laser pulse our method could be used to track a vibrational wave packet or the dissociation of a molecule \cite{vibration}. Another feature which has not been incorporated
in our model is the dependence of the ionization potential on the
internuclear distance. In fact, we have taken $E_{0}$ to be
constant, whereas, in reality, it decreases with $R$ \cite{KBK98}.
This feature, however, will only cause an overall energy shift in
the spectra. Therefore, it will not affect the distance between two
consecutive minima or maxima in the patterns for constant
internuclear distance (Figs. 6 and 7). Therefore, we expect our
method to be applicable to real physical systems and a wide
parameter range.

\begin{acknowledgments}
C.F.M.F. would like to thank L.E. Chipperfield, R. Torres, and J.P. Marangos
for useful discussions and the UK Engineering and Physical Sciences Research
Council (Advanced Fellowship, grant No. EP/D07309X/1) for financial support
\end{acknowledgments}


\begin{thebibliography}{99}
\bibitem{footnote} The SFA consists in neglecting the atomic binding
potentials when the electron is in the continuum, the external laser field
when the electron is bound, and the internal structure of the molecule.
\bibitem{footnIp} In reality, the ionization potential of a 
molecule decreases with increasing internuclear distance. This effect,
however, will only cause an overall shift in the interference patterns.
Since it will neither modify their shapes, nor the energy difference between
neighboring maxima, it is not relevant to the present discussion. For the
specific computation of this shift within the context of a zero-range model
potential see, e.g., Ref. \cite{KBK98}.
\end{thebibliography}
\end{document}